\documentclass[11pt]{article}
\usepackage{moriond,epsfig}

\def\Journal#1#2#3#4{{#1} {\bf #2}, #3 (#4)}

\def\NPA{Nucl.\ Phys.\ A}
\def\NPB{Nucl.\ Phys.\ B}
\def\PLB{Phys.\  Lett.\ B}
\def\PRL{Phys.\ Rev.\ Lett.}
\def\PRD{Phys.\ Rev.\ D}
\def\PRC{Phys.\ Rev.\ C}

\def\xbj{x_{Bj}}

\def\be{\begin{equation}}
\def\ee{\end{equation}}

\begin{document}
\vspace*{4cm}
\title{FORWARD PION PRODUCTION IN p+p AND d+Au COLLISIONS 
AT STAR\,\footnote{Presented at the XXXX$^{th}$ Rencontres de 
Moriond on QCD and High-Energy Hadronic Interactions}}

\author{ G.~RAKNESS, for the STAR Collaboration }

\address{Penn State University/Brookhaven National Lab}

\maketitle\abstracts{
Measurements are reported of the production of high energy $\pi^0$ 
mesons from the STAR experiment in p+p and d+Au collisions 
at $\sqrt{s_{NN}}=200\,$GeV and $\langle\eta\rangle=4.00$ (d beam 
direction).
The inclusive yield agrees with perturbative QCD calculations in 
p+p collisions, but is reduced in d+Au collisions.
The azimuthal correlations of the forward $\pi^0$ with charged 
hadrons at midrapidity agree with PYTHIA in p+p collisions, but
are suppressed in d+Au collisions.
The results are consistent with the conjecture that the gluon 
density in nuclei is suppressed.
}

%%%%%%%%%%%%%%%%%%%%%%%%%%%%%%%%%%%%%%%%%%%%%%%%%%%%%%%%%%%%%%%%%%%
\section{Introduction}
%%%%%%%%%%%%%%%%%%%%%%%%%%%%%%%%%%%%%%%%%%%%%%%%%%%%%%%%%%%%%%%%%%%
In contrast to the nucleon, very little is known about the density
of gluons in nuclei.\cite{hirai}
For protons, the gluon parton distribution function (PDF) is 
constrained primarily by scaling violations in deeply-inelastic 
lepton scattering (DIS) measured at the HERA collider.\cite{hera} 
The data are accurately described by QCD evolution equations that 
allow the determination of the gluon PDF.
It is found that the gluon PDF increases as the momentum fraction 
of the parton ($\xbj$) decreases.
%A general expectation is that at extremely large energies, the 
%density of gluons must be curbed by unitarity.
In nuclei, the density of gluons per unit transverse area is 
expected to be larger than in nucleons.
The interplay between large gluon densities and unitarity 
requirements on cross sections makes the nucleus a natural 
environment in which to establish if, and under which conditions, 
gluon saturation occurs.
Quantifying gluon saturation is important because of the 
expectation that the nuclear matter created in central collisions 
of Au-Au nuclei evolves from an initial state produced by the 
collisions of the low-$\xbj$ fields in each 
nucleus.\cite{gyulassymclerran}
Fixed target nuclear DIS experiments have reported a suppression of 
the inclusive structure function normalized to proton DIS at 
low-$\xbj$,\cite{fixedtarget} but are limited in the kinematic 
range necessary to determine the nuclear gluon PDF.

Many models exist which attempt to describe nuclear effects at 
low-$\xbj$.
Saturation models\,\cite{saturation} include a QCD based theory 
called the Color Glass Condensate (CGC), where gluon splitting 
is balanced by gluon-gluon recombination to create dense 
macroscopic fields.\cite{cgc}
Another approach models quarks scattering coherently from multiple 
nucleons, leading to an effective shift in the $\xbj$ 
probed.\cite{coherent}
Shadowing models suppress the nuclear gluon PDF in a standard 
factorization framework.\cite{shadowing}
Parton recombination models modify the fragmentation of a quark 
passing through many gluons.\cite{recombination}
Other descriptions include factorization breaking in heavy 
nuclei.\cite{factorization}
With the data presently available, the mechanism by which 
low-$\xbj$ gluon suppression occurs is unconstrained.

Using factorization in a perturbative QCD (pQCD) framework, the 
PDF's and fragmentation functions (FF's) measured in 
electromagnetic processes can be used in the description of 
hadronic scattering processes.
In p+p collisions at $\sqrt{s}=200\,$GeV, factorized leading 
twist pQCD calculations have been shown to quantitatively describe
inclusive $\pi^0$ production over a broad rapidity 
window.\cite{STARpi0,PHENIXpi0}
In pQCD, forward $\pi$ production in p+p collisions can be 
viewed as a probe of low-$\xbj$ gluons (g) in one proton with the 
valence quarks (q) of the other.
Recently, the yield of forward negatively charged hadrons 
($h^-$) in the d-beam direction of d+Au collisions was found to 
be reduced when normalized to p+p collisions.\cite{BRAHMS}
The reduction is especially significant since isospin effects 
are expected to reduce $h^-$ production in p+p collisions, but not 
in d+Au collisions.\cite{gsv}
The $h^-$ suppression in d+Au has been interpreted as gluon 
suppression at low-$\xbj$.

Insight on the particle production mechanism can be gained by 
analyzing the angular correlations of the forward hadron with
a coincident hadron at midrapidity.\cite{hardprobes}
A pQCD calculation of the parton kinematics probed in inclusive
forward hadroproduction reveals a broad distribution in 
$\xbj$.\cite{gsv}
Assuming collinear elastic parton ($2\rightarrow 2$) scattering, 
the pseudorapidity ($\eta=-\ln[\tan(\theta/2)]$) of the second 
particle is correlated with $\xbj$ of the probed gluon, and a 
strong back-to-back azimuthal correlation is expected.
In a saturation picture, the quark is expected to undergo 
multiple interactions through the dense gluon field, resulting 
in multiple recoil partons instead of a single recoil 
parton.\cite{coherent,monojet}

%%%%%%%%%%%%%%%%%%%%%%%%%%%%%%%%%%%%%%%%%%%%%%%%%%%%%%%%%%%%%%%%%%%
\section{Experimental Results}
%%%%%%%%%%%%%%%%%%%%%%%%%%%%%%%%%%%%%%%%%%%%%%%%%%%%%%%%%%%%%%%%%%%
The Solenoidal Tracker At RHIC (STAR) is a multipurpose detector at
BNL.
One of its principal components is a time projection chamber used for 
tracking charged particles produced at $|\eta|<1.2$.
A forward $\pi^0$ detector (FPD) comprising $7\times 7$ matrices of
$3.8\times 3.8\times 45\,$cm$^3$ Pb-glass detectors was installed to 
detect high energy $\pi^0$ mesons with $3.3< \eta < 4.1$.
Data were collected over two years of RHIC operations at 
$\sqrt{s_{NN}}=200\,$GeV.
In the 2002 run, p+p collisions were studied with a prototype
FPD.\cite{STARpi0}
In the 2003 run, p+p collisions were studied and exploratory
measurements were performed with d+Au collisions.

\begin{figure}
\begin{center}
\epsfig{figure=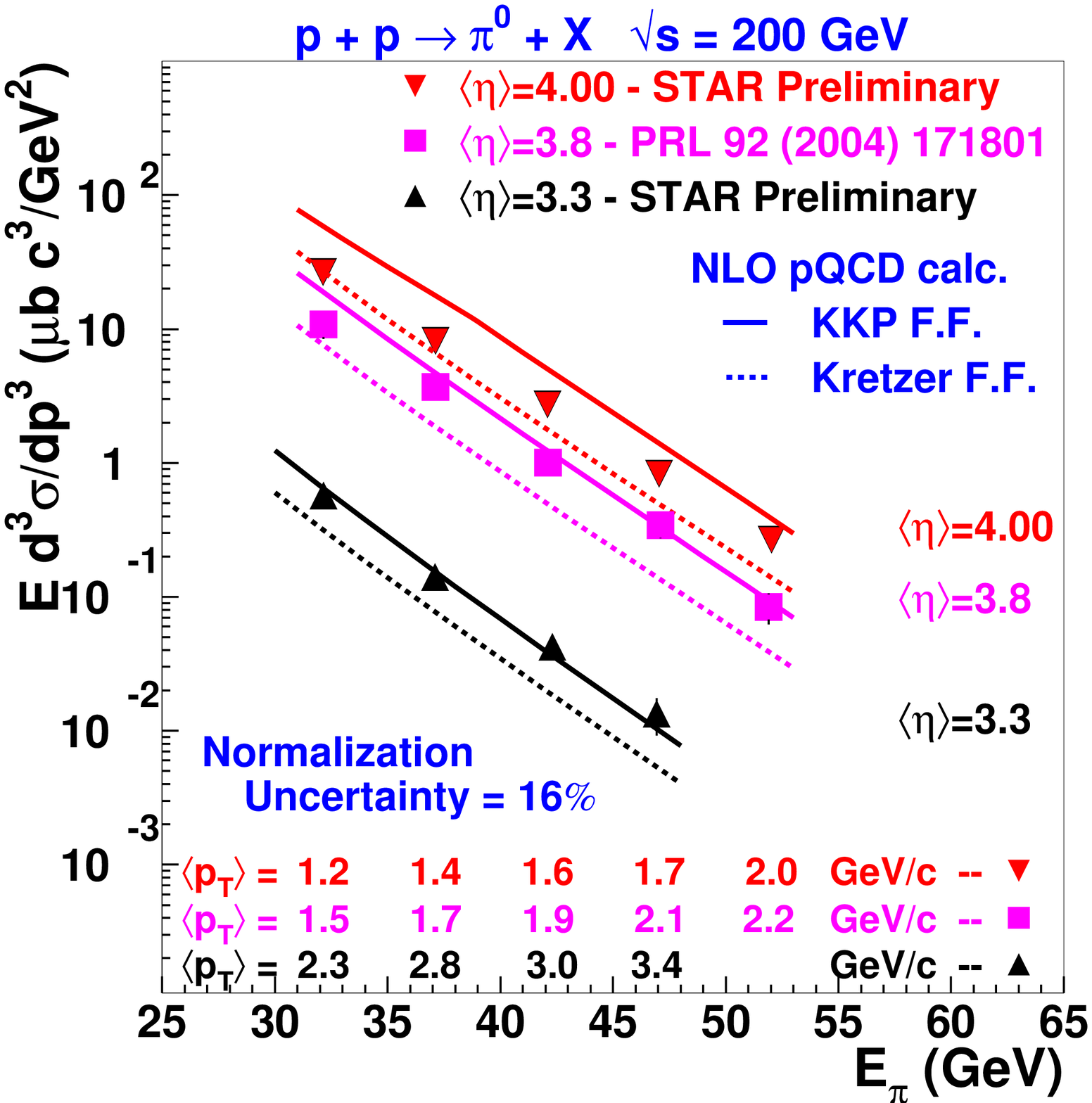,height=2.7in}
\epsfig{figure=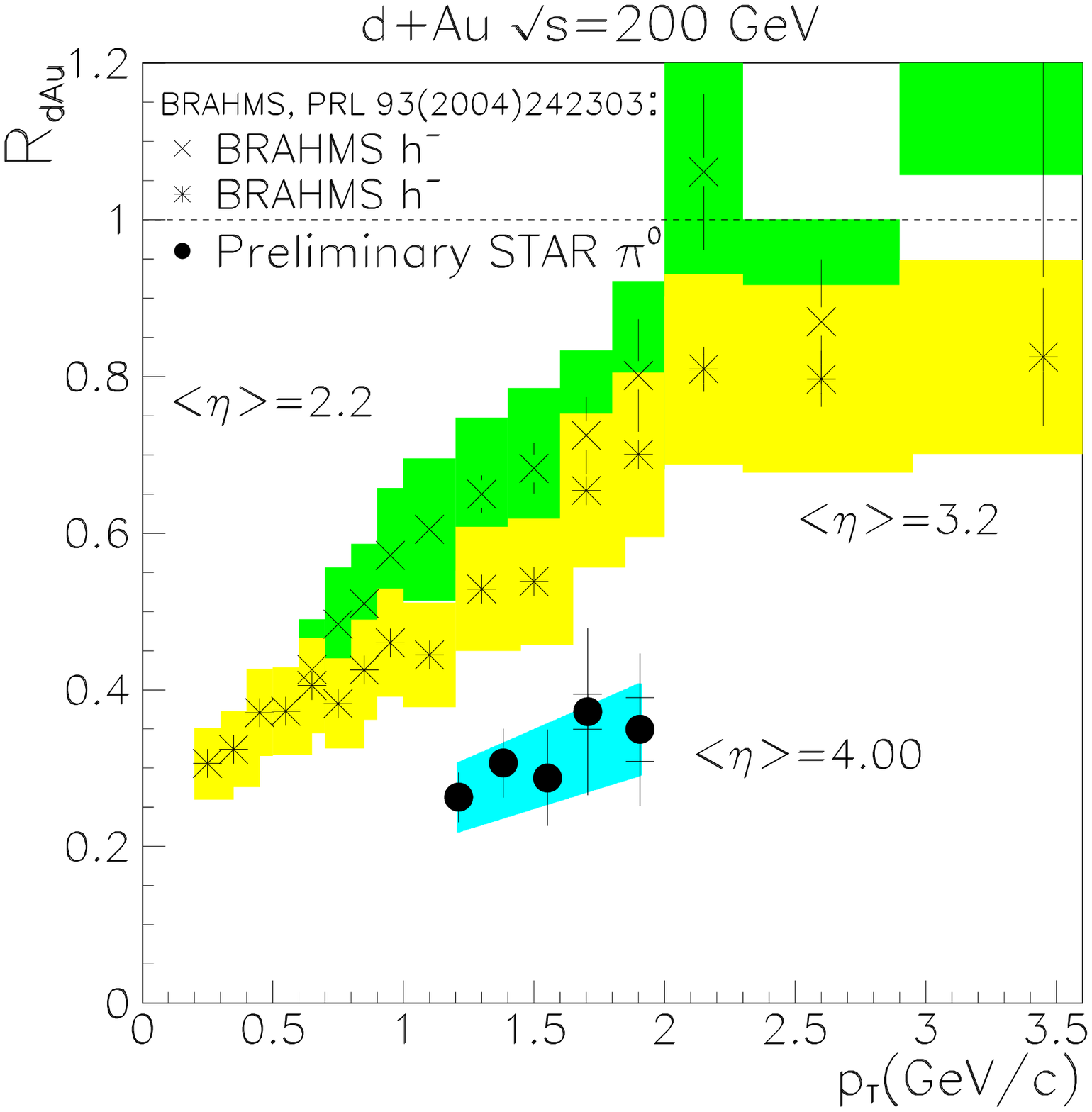,height=2.7in}
\end{center}
\caption{
Inclusive $\pi^0$ yield for p+p (left) and d+Au collisions 
normalized by p+p (right).
The pion energy ($E_\pi$) is correlated with the transverse 
momentum ($p_T$), as the FPD was at fixed values of 
pseudorapidity ($\eta$).
The inner error bars are statistical, while the outer combine 
these with the $E_\pi$- ($p_T$-) dependent systematic errors,
and are often smaller than the points.
The curves (left) are NLO pQCD calculations evaluated at fixed 
$\eta$, using different fragmentation functions.
The x's and stars (right) are BRAHMS data for $h^-$ 
production at smaller $\eta$.
\label{fig:inclusive}}
\end{figure}

The cross sections for $p+p\rightarrow\pi^0+X$ at 
$\langle\eta\rangle =3.3$\,\cite{eta33} and 
$\langle\eta\rangle = 3.8$\,\cite{STARpi0} were reported previously.
Preliminary results at $\langle\eta\rangle =4.00$ are compared 
with NLO pQCD calculations in Fig.~\ref{fig:inclusive} 
(left).\cite{dmitry}
The largest $E_\pi$-dependent systematic error at 
$\langle\eta\rangle =4.00$ is the energy calibration, resulting
in $0.08<\delta\sigma_\pi/\sigma_\pi<0.12$. 
The absolute angle uncertainty is the largest contribution to 
the normalization error.\cite{diff04}
%The calculations are performed using the CTEQ6M set of PDF's, 
%with equal renormalization and factorization scales of $p_T$.
The two curves use two sets of FF's, differing primarily in 
the gluon-to-pion FF, $D^\pi_g$.
As $p_T$ decreases, the data make the transition from 
consistency with KKP to consistency with Kretzer.
This trend is also observed at midrapidity.\cite{PHENIXpi0}
At low $p_T$, the dominant contributions to $\pi^0$ production 
are qg and gg scatterings, making $D^\pi_g$ the 
dominant FF.\cite{kretzer}

Nuclear effects on particle production are quantified by 
$R^X_{\rm dAu}$, the ratio of the inclusive yield of $X$ in d+Au 
compared to p+p collisions normalized by the number of 
nucleon-nucleon collisions.
This same measure was used at midrapidity, where it was found
$R^{\,h^\pm}_{\rm dAu}\stackrel{>}{_\sim}1$.\cite{STARdau}
The solid circles in Fig.~\ref{fig:inclusive} (right) are STAR
preliminary data for $R^{\,\pi^0}_{\rm dAu}$ at 
$\langle\eta\rangle=4.00$.
On average, 0.5 more photons per event are observed in d+Au 
collisions than in p+p collisions for events with $>30\,$GeV 
detected in the FPD.
This leads to the largest $p_T$-dependent systematic error which 
comes from the background correction to the d+Au yield.
The ratio $R^{\,\pi^0}_{\rm dAu}$ at $\langle\eta\rangle =4.00$ is 
significantly smaller than $R^{\,h^-}_{\rm dAu}$ at smaller 
$\eta$,\cite{BRAHMS} consistent with the trend expected from models 
which suppress the nuclear gluon 
density.\cite{cgc,coherent,shadowing,recombination,factorization}
Linearly extrapolating $R^{\,h^-}_{\rm dAu}$ to $\eta =4$,
$R^{\,\pi^0}_{\rm dAu}$ is systematically smaller, consistent with 
expectations of isospin suppression of 
$p+p\rightarrow h^-+X$.\cite{gsv}

\begin{figure}
\epsfig{figure=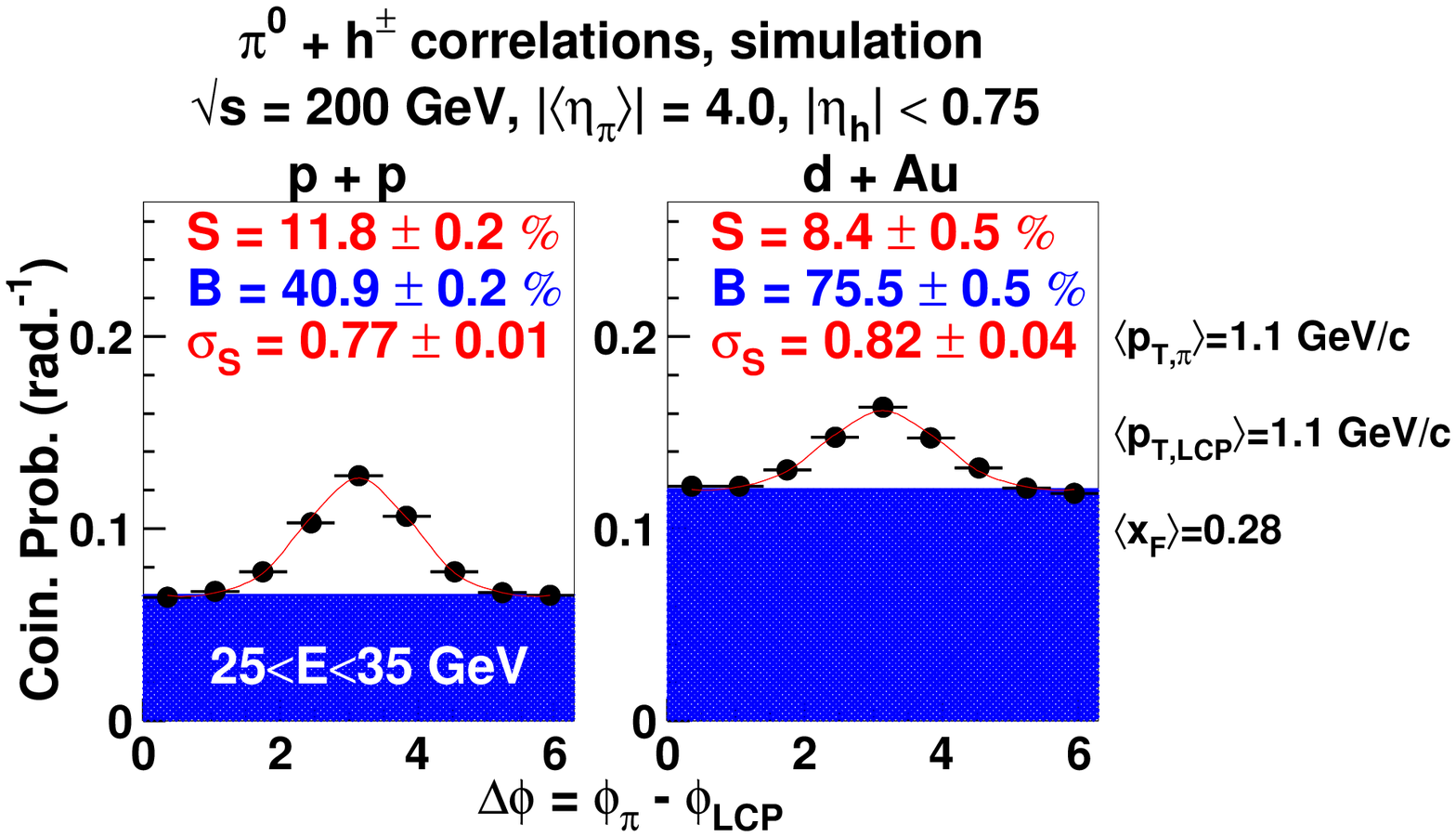,height=1.9in}
\epsfig{figure=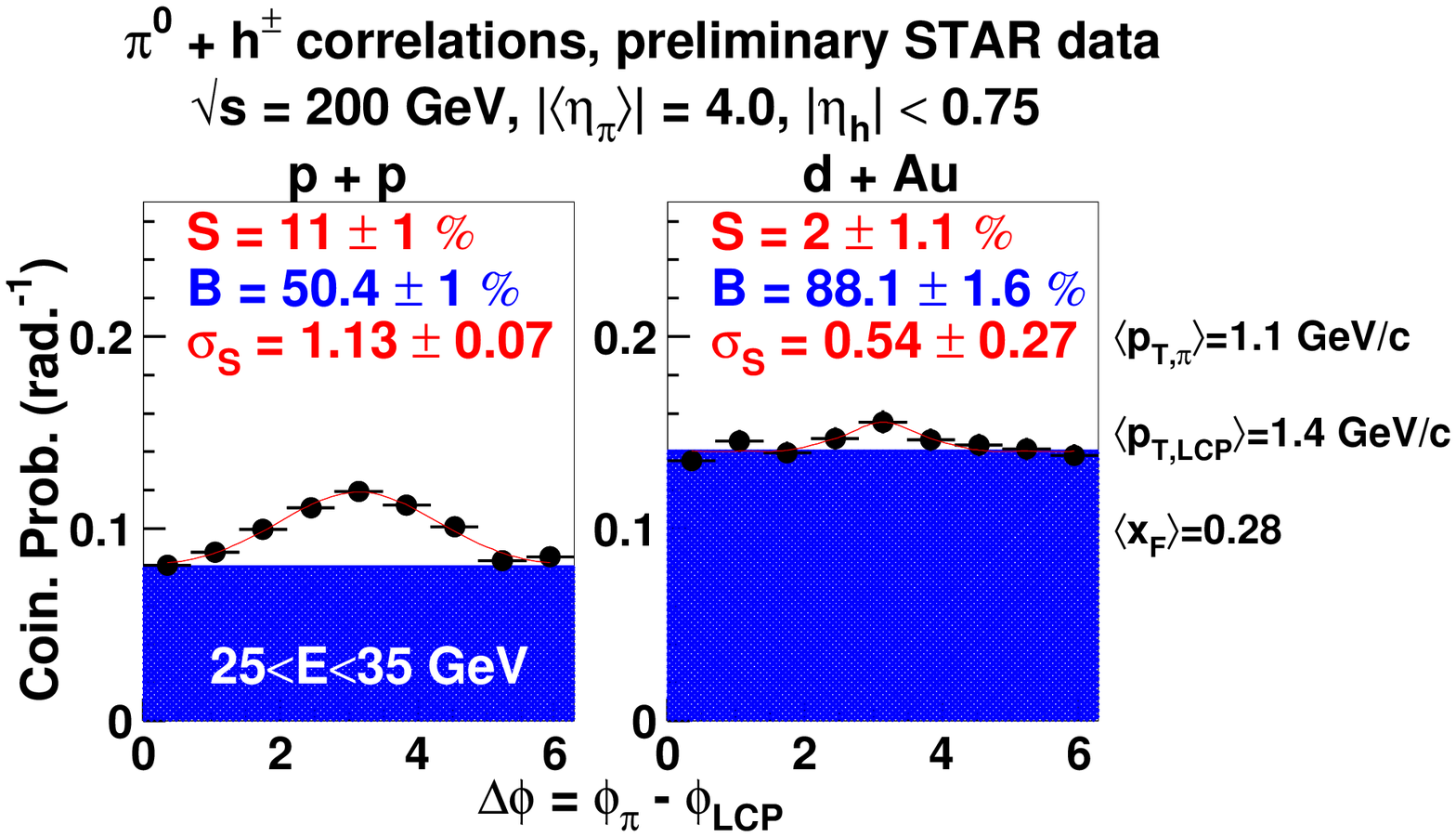,height=1.9in}
\caption{
Coincidence probability as a function of azimuthal angle difference 
between the forward $\pi^0$ meson and the leading charged particle 
at midrapidity.
The left two distributions are simulations, using PYTHIA for p+p 
and HIJING for d+Au collisions.
The right two plots are data.
The curves are fits described in the text.
\label{fig:azimuthal}}
\end{figure}

%%%%%%%%%%%%%%%%%%%%%%%%%%%%%%%%%%%%%%%%%%%%%%%%%%%%%%%%%%%%%%%%%%%
%\subsection{Di-Hadron Azimuthal Correlations}
%%%%%%%%%%%%%%%%%%%%%%%%%%%%%%%%%%%%%%%%%%%%%%%%%%%%%%%%%%%%%%%%%%%
Exploratory measurements of the azimuthal correlations between a
forward $\pi^0$ and midrapidity $h^\pm$ were completed for p+p and 
d+Au collisions.
The data analysis and detailed comparisons to simulations were
presented earlier.\cite{dis2004}
The left two plots in Fig.~\ref{fig:azimuthal} are simulations,
%including detector resolution and reconstruction efficiency, 
using PYTHIA 6.222\,\cite{pythia} for p+p and HIJING 
1.381\,\cite{hijing} for d+Au collisions.
The right two plots in Fig.~\ref{fig:azimuthal} are 
data.
%The efficiency uncorrected average multiplicity of charged particles
%with $|\eta|<0.75$ and $p_T >0.2\,$GeV/c for events coincident with a
%forward $\pi^0$ is 5.1 for p+p, and 14.4 for d+Au.
%This multiplicity is approximately 10\% larger than what is observed
%for minimum-bias events.
The leading charged particle (LCP) analysis selects the midrapidity
track with the highest $p_T>0.5\,$GeV/c, and computes the azimuthal 
angle difference $\Delta\phi=\phi_{\pi^0}-\phi_{LCP}$ for each event.
%The $\Delta\phi$ distributions are normalized by the number of 
%$\pi^0$ observed at $\langle \eta_{\pi} \rangle=4.0$.
The normalized distributions are fit with the sum of a constant and a 
Gaussian centered at $\pi$.
Correlations near $\Delta\phi=0$ are not expected because of 
the large $\eta$ separation between the $\pi^0$ and the LCP.
The fit values are the areas under both the back-to-back peak (S) 
and the constant (B), and the width of the peak ($\sigma_S$). 
%having contributions from transverse momentum in the hadronization 
%of the jets and the momentum imbalance between the pair of jets 
%($k_T$).
%The errors on the fitted parameters are based on the full error 
%matrix.

PYTHIA reproduces most features of the p+p data.
The back-to-back peak arises from $2\rightarrow 2$ scattering, 
resulting in forward and midrapidity jets that fragment into the 
$\pi^0$ and LCP, respectively.
The width of the peak is smaller in PYTHIA than in the data, 
indicating that the momentum imbalance between the jets is too 
small in PYTHIA.
This was also seen for back-to-back jets at the Tevatron.\cite{d0}
%A component of B comes from $2\rightarrow 3$ partonic processes, 
%fully accounted for in NLO pQCD calculations, and approximated by 
%parton showers in PYTHIA.

%Comparing the d+Au data with the simulations and the p+p data,
%We observe a large increase of B$_{\rm dAu}$ relative to 
%B$_{\rm pp}$.
%HIJING models d+Au collisions using PYTHIA for inelastic
%nucleon-nucleon interactions and the Glauber model to account for
%multiple collisions.
%From HIJING, the growth in B arises from additional nucleon-nucleon 
%collisions as the deuteron and Au beams interact.
%HIJING does not predict a significant difference between the width 
%$\sigma^{\rm dAu}$ relative to $\sigma^{\rm pp}$.
%For the data, $\sigma_S^{\rm dAu}$ is much smaller than
%$\sigma_S^{\rm pp}$, most likely reflecting the inadequacy of the
%functional form used to represent $\Delta\phi_{\rm dAu}$.
The back-to-back peak is significantly smaller in d+Au collisions 
than in p+p, qualitatively consistent with the coherent 
scattering\,\cite{coherent} and CGC\,\cite{monojet} models.
%Part of the reduction is due to the fact that S is proportional to
%$\sigma$ and the fit results in a width for $\Delta\phi_{\rm dAu}$
%that is likely to be unphysically small.
HIJING includes a model of shadowing for nuclear PDF's, and predicts 
a sizable back-to-back peak.
This is not observed in the preliminary STAR data.
Complete assessment of systematic errors is underway.
 
%%%%%%%%%%%%%%%%%%%%%%%%%%%%%%%%%%%%%%%%%%%%%%%%%%%%%%%%%%%%%%%%%%%
%\section{Summary}
%%%%%%%%%%%%%%%%%%%%%%%%%%%%%%%%%%%%%%%%%%%%%%%%%%%%%%%%%%%%%%%%%%%
In summary, inclusive forward $\pi^0$ cross sections from p+p 
collisions at $\sqrt{s}=200\,$GeV are consistent with NLO pQCD 
calculations.
Azimuthal correlations of the forward $\pi^0$ with midrapidity 
$h^\pm$ are described by PYTHIA, which uses leading order pQCD 
with parton showers.
The success of both of these calculations implies that forward 
$\pi^0$ production arises from partonic scattering in p+p 
collisions at $\sqrt{s}=200\,$GeV.
In d+Au collisions, the inclusive yield of forward $\pi^0$ mesons 
is found to be significantly reduced compared to p+p.
The $\eta$ dependence of the reduction is consistent with models 
which suppress the gluon density in nuclei, in addition to 
exhibiting isospin effects at these kinematics.
Exploratory studies suggest that the back-to-back peak of forward 
$\pi^0$ mesons with midrapidity $h^\pm$ is suppressed in d+Au
relative to p+p collisions.
This is qualitatively consistent with expectations that the particle 
production in a dense gluon medium differs from conventional
leading-twist NLO pQCD expectations.
More data for forward particle production and di-hadron correlations
in d+Au collisions are required to reach a definitive conclusion
about the existence of gluon saturation in the Au nucleus,
or if an alternative description of the forward rapidity suppression 
is valid.
A quantitative theoretical understanding of the rapidity and $p_T$
dependence of di-hadron correlations would facilitate experimental
tests of a possible color glass condensate.

%%%%%%%%%%%%%%%%%%%%%%%%%%%%%%%%%%%%%%%%%%%%%%%%%%%%%%%%%%%%%%%%%%%
\section*{References}
%%%%%%%%%%%%%%%%%%%%%%%%%%%%%%%%%%%%%%%%%%%%%%%%%%%%%%%%%%%%%%%%%%%

\end{document}